\documentclass[12pt]{article}
\usepackage[a4paper,total={18cm,27cm}]{geometry}
\usepackage{graphicx}
\usepackage{authblk}
\usepackage{amsmath}
\usepackage[utf8]{inputenc}
\usepackage{hyperref}
\usepackage{verbatim}
\usepackage{cite}

\begin{document}


\title{Dynamical Phases and Resonance Phenomena\\ in Information-Processing Recurrent Neural Networks}


\author[1]{Claus Metzner}
\author[1,2,3]{Patrick Krauss}
\affil[1]{\small Neuroscience Lab, University Hospital Erlangen, Germany}
\affil[2]{\small Cognitive Computational Neuroscience Group, Friedrich-Alexander-University Erlangen-Nuremberg, Germany}
\affil[3]{\small Pattern Recognition Lab, Friedrich-Alexander-University Erlangen-Nuremberg, Germany}
\maketitle


\begin{abstract}\large
Recurrent neural networks (RNNs) are complex dynamical systems, capable of ongoing activity without any driving input. The long-term behavior of free-running RNNs, described by periodic, chaotic and fixed point attractors, is controlled by the statistics of the neural connection weights, such as the density $d$ of non-zero connections, or the balance $b$ between excitatory and inhibitory connections. However, for information processing purposes, RNNs need to receive external input signals, and it is not clear which of the dynamical regimes is optimal for this information import. We use both the average correlations $C$ and the mutual information $I$ between the momentary input vector and the next system state vector as quantitative measures of information import and analyze their dependence on the balance and density of the network. Remarkably, both resulting phase diagrams $C(b,d)$ and $I(b,d)$ are highly consistent, pointing to a link between the dynamical systems and the information-processing approach to complex systems. Information import is maximal not at the 'edge of chaos', which is optimally suited for computation, but surprisingly in the low-density chaotic regime and at the border between the chaotic and fixed point regime. Moreover, we find a completely new type of resonance phenomenon, called 'Import Resonance' (IR), where the information import shows a maximum, i.e. a peak-like dependence on the coupling strength between the RNN and its input. IR complements Recurrence Resonance (RR), where correlation and mutual information of successive system states peak for a certain amplitude of noise added to the system. Both IR and RR can be exploited to optimize information processing in artificial neural networks and might also play a crucial role in biological neural systems.  

\end{abstract}




\clearpage
\section*{Introduction}

At present, the field of Machine Learning is strongly dominated by feed-forward neural networks, which can be optimized to approximate an arbitrary vectorial function $\mathbf{y} = \mathbf{f}(\mathbf{x})$ between the input and output spaces \cite{hornik1989multilayer, cybenko1992approximation, funahashi1989approximate}. Recurrent neural networks (RNNs) however are a much broader class of models, which encompass the feed-forward architectures as a special case, but which also include partly recurrent systems, such as contemporary LSTMs (long short-term memories) \cite{hochreiter1997long} and classical Jordan or Elman networks \cite{cruse2006neural}, up to fully connected systems without any layered structure, such as Hopfield networks \cite{ilopfield1982neural} or Boltzmann machines \cite{hinton1983optimal}. Due to the feedback built into these systems, RNNs can learn robust representations \cite{farrell2019recurrent}, and are ideally suited to process sequences of data such as natural language \cite{lecun2015deep, schilling2021quantifying}, or to perform sequential-decision tasks such as spatial navigation \cite{banino2018vector, gerum2020sparsity}. Furthermore, RNNs can act as autonomous dynamical systems that continuously update their internal state $\mathbf{s}_t$ even without any external input \cite{gros2009cognitive}, but it is equally possible to modulate this internal dynamics by feeding in external input signals $\mathbf{x}_t$ \cite{jaeger2014controlling}. Indeed, it has been shown that RNNs can approximate any open dynamical system $\mathbf{s}_{t+1} = \mathbf{g}(\mathbf{s}_t,\mathbf{x}_t))$ to arbitrary precision \cite{schafer2006recurrent}.

It is therefore not very surprising that biological neural networks are also highly recurrent in their connectivity \cite{squire2012fundamental,binzegger2004quantitative} making RNNs to versatile tools of neuroscience research \cite{barak2017recurrent, maheswaranathan2019universality}. Modelling natural RNNs in a realistic way requires the use of probabilistic, spiking neurons, but even simpler models with deterministic neurons already have highly complex dynamical properties and offer fascinating insights into how structure controls function in non-linear systems \cite{krauss2019analysis,krauss2019weight}. For example, we have demonstrated that by adjusting the density $d$ of non-zero connections and the balance $b$ between excitatory and inhibitory connections in the RNN's weight matrix, it is possible to control whether the system will predominantly end up in a periodic, chaotic, or fixed point attractor \cite{krauss2019weight}. Understanding and controlling the behavior of RNNs is of crucial importance for practical applications \cite{haviv2019understanding}, especially as meaningful computation, or information processing, is believed to be only possible at the 'edge of chaos' \cite{bertschinger2004real,natschlager2005edge,legenstein2007edge,schrauwen2009computational,busing2010connectivity,toyoizumi2011beyond,dambre2012information}.

In this paper, we continue our investigation of RNNs with deterministic neurons and random, but statistically controlled weight matrices. Yet, the present work focuses on another crucial precondition for practical RNN applications: the ability of the system to 'take up' external information and to incorporate it into the ongoing evolution of the internal system states. For this purpose, we first set up quantitative measures of information import, in particular $C(\mathbf{x}_t,\mathbf{s}_{t+1})$, the RMS average of all pairwise neural correlations between the momentary input $\mathbf{x}_t$ and the subsequent system state $\mathbf{s}_{t+1}$, as well as $I(\mathbf{x}_t,\mathbf{s}_{t+1})$, an approximation for the mean pairwise mutual information between the same two quantities. We then compute these measures for all possible combinations of the structural parameters $b$ (balance) and $d$ (density) on a grid, resulting in high-resolution phase diagrams $C(b,d)$ and $I(b,d)$. This reveals that the regions of phase space in which information processing (computation) and information import (representation) are optimal, surprisingly do not coincide, but nevertheless have a small area of phase space in common. We speculate that this overlap region, where both crucial functions are simultaneously possible, may represent a 'sweet spot' for practical RNN applications and might therefore be exploited by biological nervous systems.


\section*{Results}

\subsection*{Free-running network}

\begin{figure}[ht!]
\centering
\includegraphics[width=1.0\linewidth]{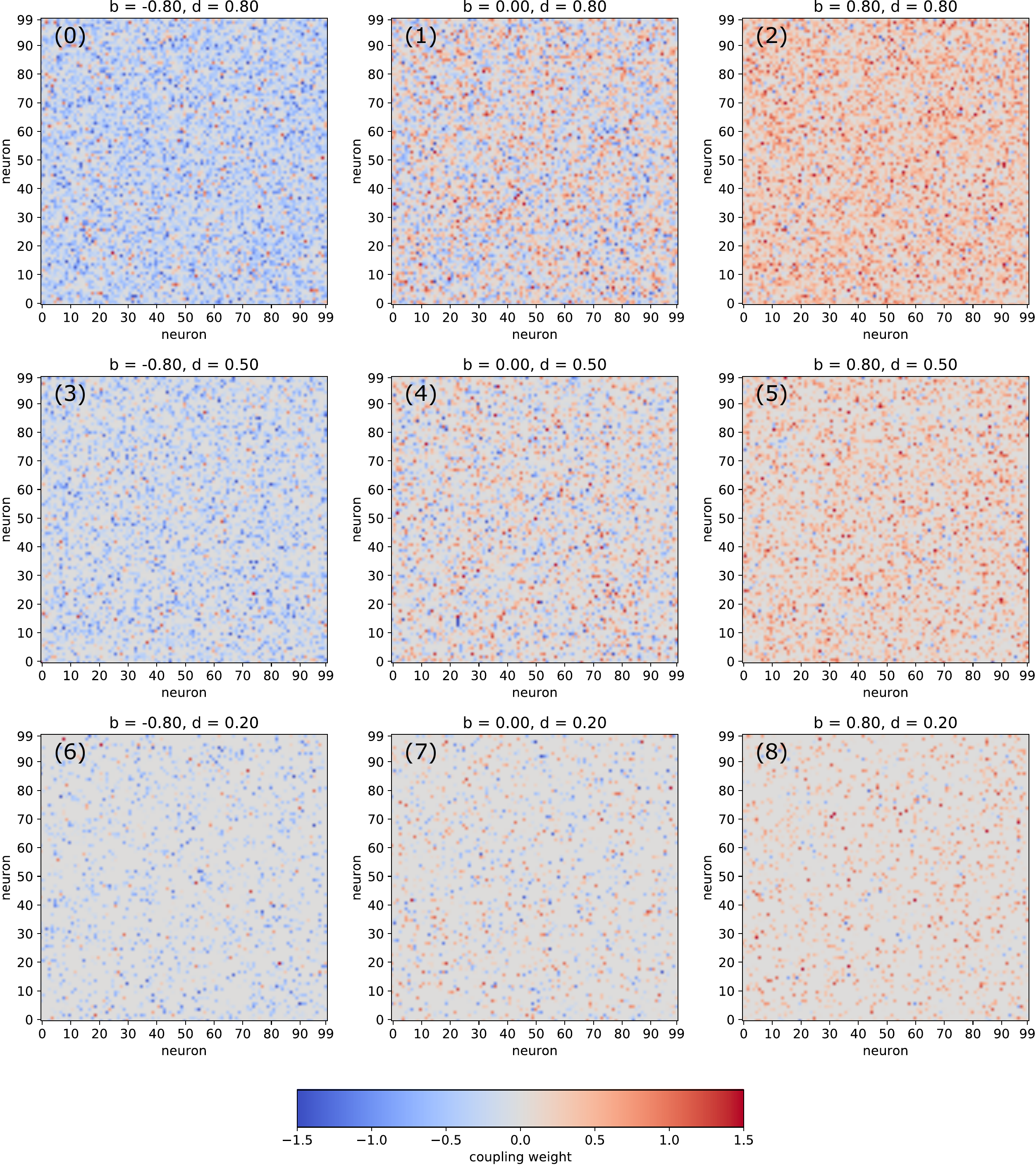}
\caption{Examples of weight matrices for selected combinations of the balance $b$ between excitatory and inhibitory connections and the density $d$ of non-zero connections in a RNN.} 
\label{figure_1}
\end{figure}

In the following, we are analyzing networks  composed of $N_{neu}=100$ deterministic neurons with arcustangent activation functions. The random matrix of connection weights is set up in a controlled way, so that the density $d$ of non-zero connections as well the balance $b$ between excitatory and inhibitory connections can be pre-defined independently (for details see Methods section). Visualizations of typical weight matrices for different combinations of the statistical control parameters $d$ and $b$ are shown in Fig. \ref{figure_1}.

\begin{figure}[ht!]
\centering
\includegraphics[width=1.0\linewidth]{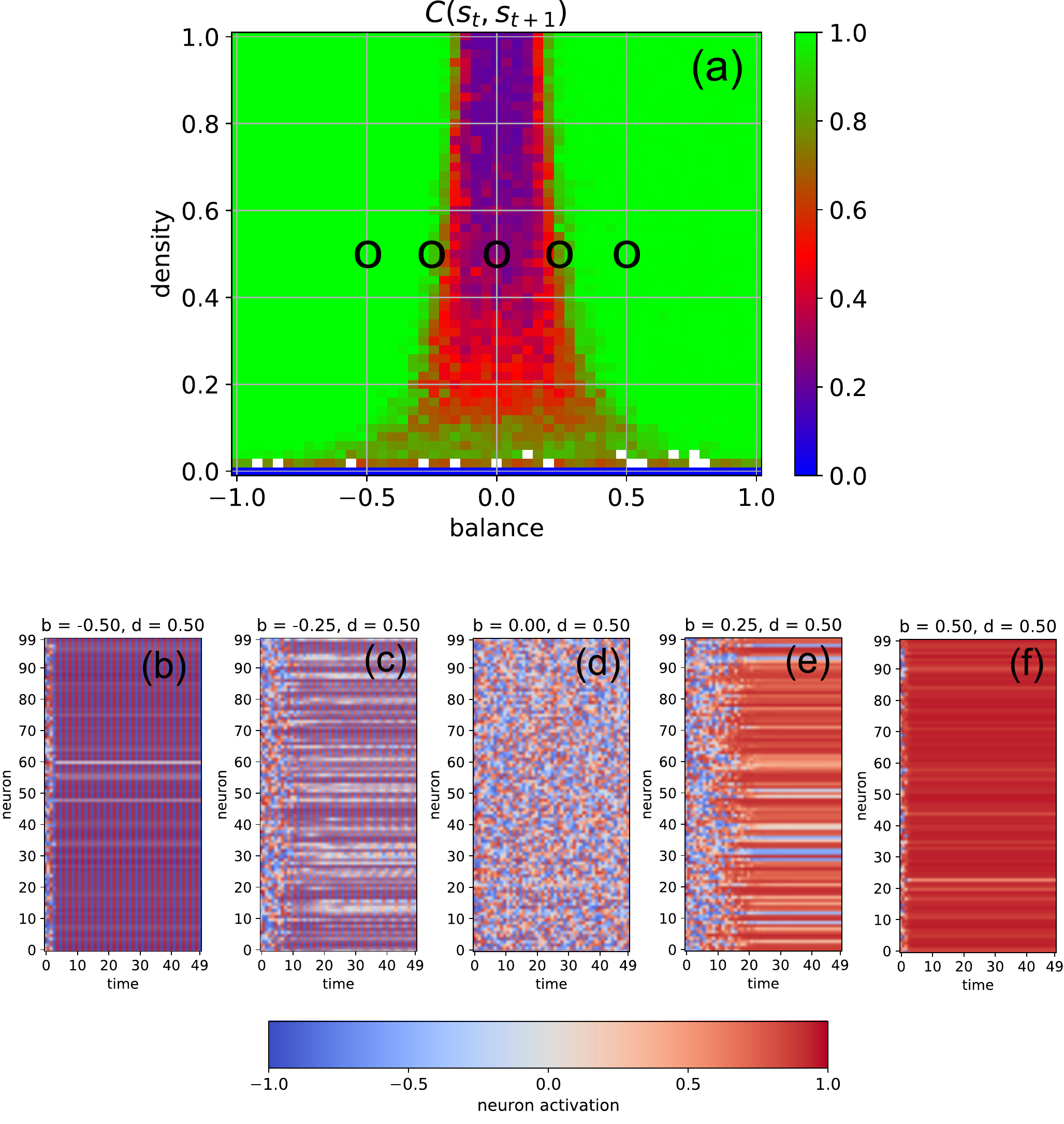}
\caption{Dynamical phases of a free-running RNN, controlled by the structural parameters $b$ (balance) and $d$ (density). (a): Phase diagram of the correlation $C(\mathbf{s}_t,\mathbf{s}_{t+1})$ between successive neuron activations, as defined in the methods section. The three basic regimes are the oscillatory phase for negative balances (large correlations), the chaotic phase for balances close to zero (small correlations), and the fixed point phase for positive balances (large correlations). (b-f): Typical time dependence of the neural activations for fixed density ($d=0.5$) and balances increasing from $b=-0.5$ to $b=+0.5$. The system behavior evolves from almost homogeneous oscillations (b), to a partly oscillatory and partly chaotic state (c), to fully chaotic behavior (d), to a state with different fixed points for each neuron (e), and finally to an almost global fixed point attractor (f). } 
\label{figure_2}
\end{figure}

We first investigate free-running networks without external input and compute a dynamical phase diagram $C_{ss}(b,d)$ of the average correlation $C_{ss}=C(\mathbf{s}_t,\mathbf{s}_{t+1})$ between subsequent system states (Fig. \ref{figure_2} a, for details see Methods section). The resulting landscape is mirror-symmetric with respect to the line $b=0$, due to the symmetric activation functions of our model neurons, combined with the definition of the balance parameter. Apart from the region of very low connection densities with $d\le 0.1$, the phase space consists of three major parts: the oscillatory regime in networks with predominantly inhibitory connections ($b\ll0$, left green area in Fig. \ref{figure_2} a), the chaotic regime with approximately balanced connections ($b\approx0$, central blue and red area in Fig. \ref{figure_2} a), and the fixed point regime with predominantly excitatory connections ($b\gg0$, right green area in Fig. \ref{figure_2} a).

It is important to note that $C(\mathbf{s}_t,\mathbf{s}_{t+1})$ is a root-mean-square (RMS) average over all the pairwise correlations between subsequent neural activations (so that negative and positive correlations are not distinguished), and that these pairwise correlations are properly normalized in the sense of a Pearson coefficient (each ranging between -1 and +1 before the RMS is computed). For this reason,  
$C(\mathbf{s}_t,\mathbf{s}_{t+1})$ is close to one (green) both in the oscillatory and in the fixed point regimes, where the system is behaving regularly. By contrast, $C(\mathbf{s}_t,\mathbf{s}_{t+1})$ is close to zero (blue) in the high-density part of the chaotic regime, where the time-evolution of the system is extremely irregular. Moreover, we find medium-level correlations (red) in the low-density part of the chaotic regime, and also at the borders between the chaotic and the two regular regimes. Since medium-level correlations are optimally suited for information processing, it is remarkable that this can take place not only at the classical 'edge of chaos' (between the oscillatory and the chaotic regime), but also in other (and less investigated) regions of the network's dynamical phase space. However, as soon as we couple the system to an external input of significant strength ($\eta=0.5$), we find that only the classical edge of chaos remains available for information processing (Fig. \ref{figure_3} a). 

In order to verify the nature of the three major dynamical regimes, we investigate the complete time evolution of the neural activations for selected combinations of the control parameters $b$ and $d$. In particular, we fix the connection density to $d=0.5$ and gradually increase the balance from $b=-0.5$ to $b=+0.5$ in five steps (Fig. \ref{figure_2} b-f). As expected, we find almost perfect oscillations (here with a period of two time steps) for $b=-0.5$ (case (b)), at least after the transient period in which the system is still carrying a memory of the random initialization of the neural activations. At $b=0$ (case (d)), we find completely irregular, chaotic behaviour, and at $b=+0.5$ (case (d)) almost all neurons reach the same fixed point. However, the cases close to the two edges of the chaotic regime reveal an interesting intermediate dynamic behavior: For $b=-0.25$ (case (c)), most neurons are synchronized in their oscillations, but some are out of phase, or even behave non-periodically. For $b=+0.25$ (case (e)), most neurons reach (approximately) the same shared fixed point, but some end up in a different, individual fixed point. 
Moreover, we observe that the memory time $\tau$ of the system for the information imprinted by the initialization (that is, the duration of the transient phase) depends systematically on the balance parameter: Deep within the oscillatory regime ($b=-0.5$, case (b)), $\tau$ is short. As we approach the chaotic regime ($b=-0.25$, case (c)), $\tau$ increases, finally becoming 'infinitely' long at $b=0$ (case (d)). Indeed, from this viewpoint the chaotic dynamics may be interpreted as the continuation of the transient phase. Finally, $\tau$ is decreasing again as we move deeper into the fixed point regime (cases (e,f)).

\subsection*{Network driven by continuous random input}

Next, we feed into the network a relatively weak external input (with a coupling strength of $\eta=0.5$), consisting of independent normally distributed random signals that are continuously injected to each of the neurons (for details see Method section).

\begin{figure}[ht!]
\centering
\includegraphics[width=1.0\linewidth]{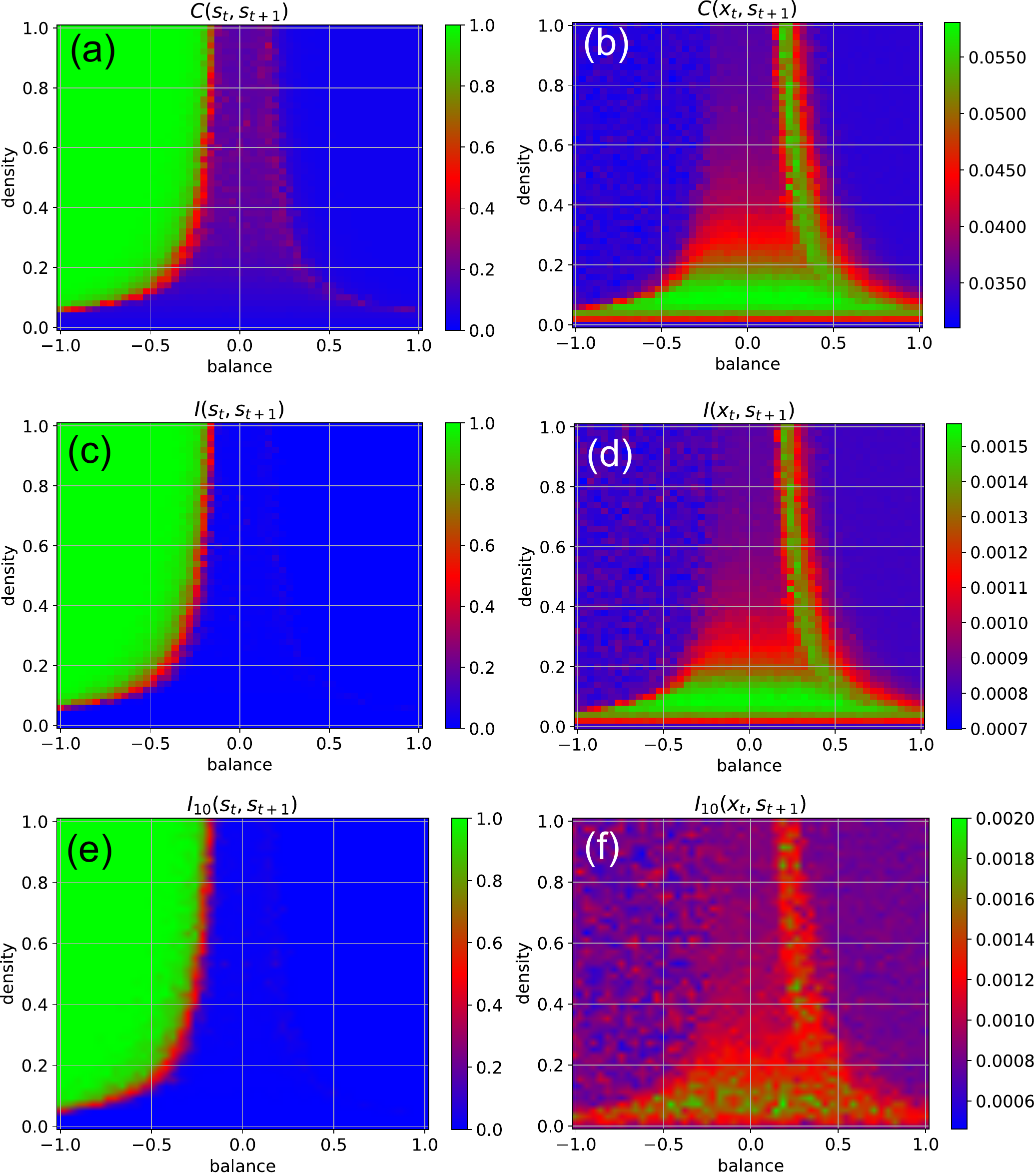}
\caption{Dynamical phases of a RNN driven by external input in the form of continuous random signals that are coupled independently to all neurons with a coupling constant of $\eta=0.5$. The suitability of the system for information processing is characterized by the statistical dependency between subsequent states (left column), the suitability for information import by the statistical dependency between the input $\mathbf{x}_t$ and the subsequent state $\mathbf{s}_{t+1}$ (right column). First row (a,b): Root-mean-square of correlations. Second row (c,d): Mean pairwise mutual information. Information import is optimal in the low-density chaotic regime and at the border between the chaotic and fixed point regime (red and green color in right column). Third row (e,f):  Approximation of the mean pairwise mutual information, where only a sub-population of 10 neurons is included to the evaluation.} 
\label{figure_3}
\end{figure}

As already mentioned in the last section, the external input destroys the medium-level state-to-state correlations  $C(\mathbf{s}_t,\mathbf{s}_{t+1})$ in most parts of the chaotic regime, except at the classical edge of chaos (Fig. \ref{figure_3} a, red). Moreover, the input also brings the correlations in the fixed point regime down to a very small value, as now the external random signals are superimposed onto the fixed points of the neurons. We can therefore verify in our model the known fact that genuine information processing cannot take place in dynamical regimes with either too strong or too weak correlations, but is instead only possible at the classical edge of chaos.

However, another practically important factor is the ability of neural networks to take up external information at any point in time and to incorporate it into their system state. We quantify this ability of information import by the RMS-averaged correlation $C(\mathbf{x}_t,\mathbf{s}_{t+1})$ between momentary input and subsequent system state. Surprisingly, we find that information import is best, i.e. $C(\mathbf{x}_t,\mathbf{s}_{t+1})$ is large, in the low-density part of the chaotic regime, including the lowest part of the classical edge of chaos (region between chaotic and oscillatory regimes), but also at the opposite border between the chaotic and fixed point regimes (Fig. \ref{figure_3} b, green and red). 
We thus come to the conclusion that (at least for weak external inputs with $\eta=0.5$) our network model is simultaneously capable of information import and information processing only in the low-density part of the classical edge of chaos.

To backup this unexpected finding, we also quantify information processing and information import by the average pair-wise state-to-state mutual information $I(\mathbf{s}_t,\mathbf{s}_{t+1})$ (Fig. \ref{figure_3} c), and the mutual information between the momentary input and the subsequent system state $I(\mathbf{x}_t,\mathbf{s}_{t+1})$ (Fig. \ref{figure_3} d), respectively. These mutual-information-based measures of information import and processing can also capture possible non-linear dependencies, but are computationally much more demanding (for details see Method section).

Remarkably, it turns out that the corresponding phase diagrams for information import ($C(\mathbf{x}_t,\mathbf{s}_{t+1})$ and $I(\mathbf{x}_t,\mathbf{s}_{t+1})$), as well as those for information processing ($C(\mathbf{s}_t,\mathbf{s}_{t+1})$ and $I(\mathbf{s}_t,\mathbf{s}_{t+1})$), are almost identical (Fig. \ref{figure_3}, compare a, b with c, d). This unexpected finding
points to a possible link bridging the gap between information-processing and dynamical approaches to complexity science \cite{mediano2021integrated}.

Furthermore, we compare the results to a computationally more tractable approximation of the mean pairwise mutual information, where only a sub-population of 10 neurons is included to the evaluation. It also shows the same basic characteristics (Fig. \ref{figure_3} e, f), implicating the possibility to approximate mutual information in large dynamical systems, where an exhaustive sampling of all joint probabilities necessary to calculate entropy and mutual information is impractical or impossible.

\subsection*{Effect of increasing coupling strength}

\begin{figure}[ht!]
\centering
\includegraphics[width=0.5\linewidth]{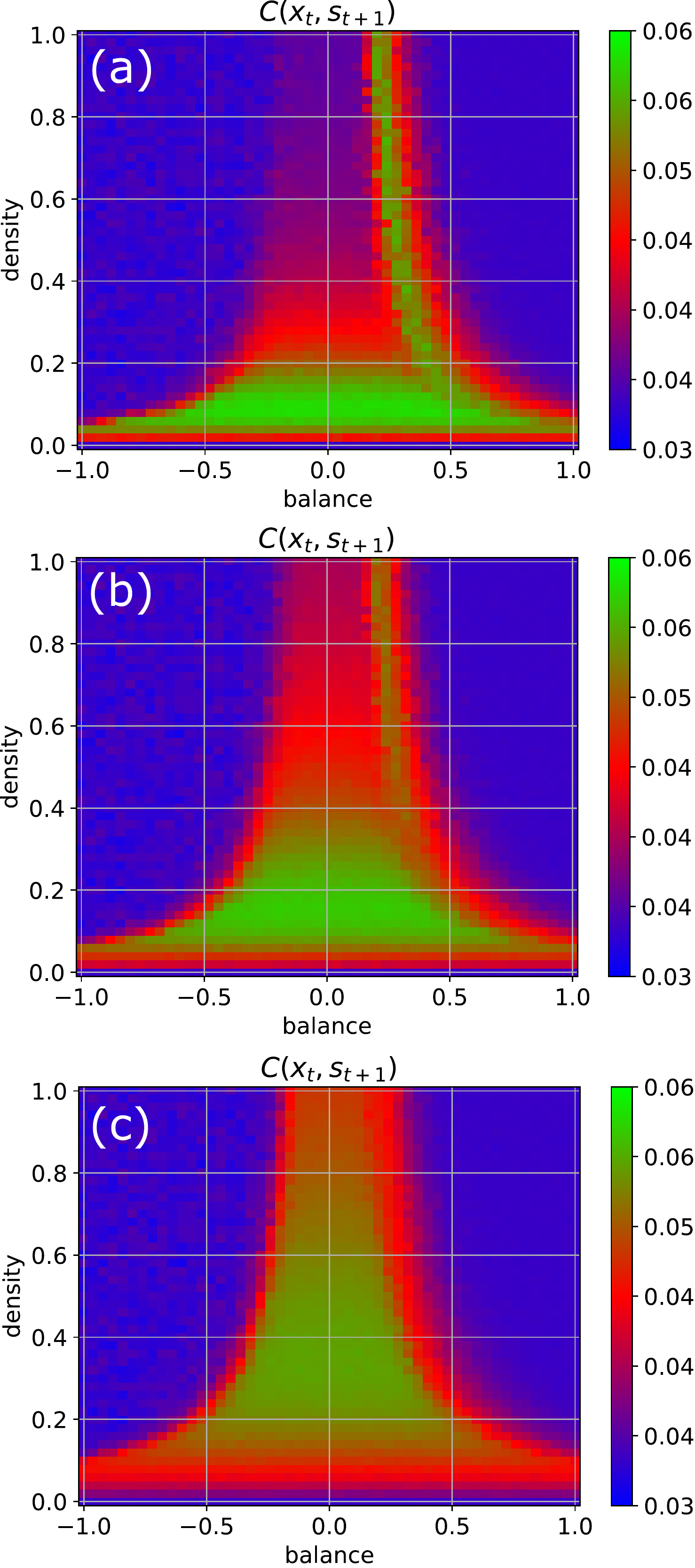}
\caption{Information import as a function of the coupling strength $\eta$ between the RNN neurons and the external input signals. For weak coupling ($\eta=0.5$ in (a)), only the low-density chaotic regime and the border between the chaotic and fixed point regime are suitable for information import. As the coupling in increases from $\eta=1$ in (b) to $\eta=2$ in (c), the correlations between input $\mathbf{x}_t$ and subsequent RNN states $\mathbf{s}_{t+1}$ become gradually large throughout the complete chaotic regime.} 
\label{figure_4}
\end{figure}

When the coupling strength to the random input signals is step-wise increased from $\eta=0.5$ to $\eta=1$ and finally to $\eta=2$ (Fig. \ref{figure_4}), we observe that also the higher density parts of the chaotic regime become eventually available for information import (green color).

\subsection*{Import Resonance (IR) and Recurrence Resonance (RR)}

Next, we increase the coupling strength $\eta$ gradually from zero to a very large value of 20, at which the random input already dominates the system dynamics. For this numerical experiment, we keep the balance and density parameters fixed at $b\!=\!0,\;d\!=\!0.5$ (chaotic regime) and $b\!=\!0.5,\;d\!=\!0.5$ (fixed point regime), respectively.

\begin{figure}[ht!]
\centering
\includegraphics[width=1\linewidth]{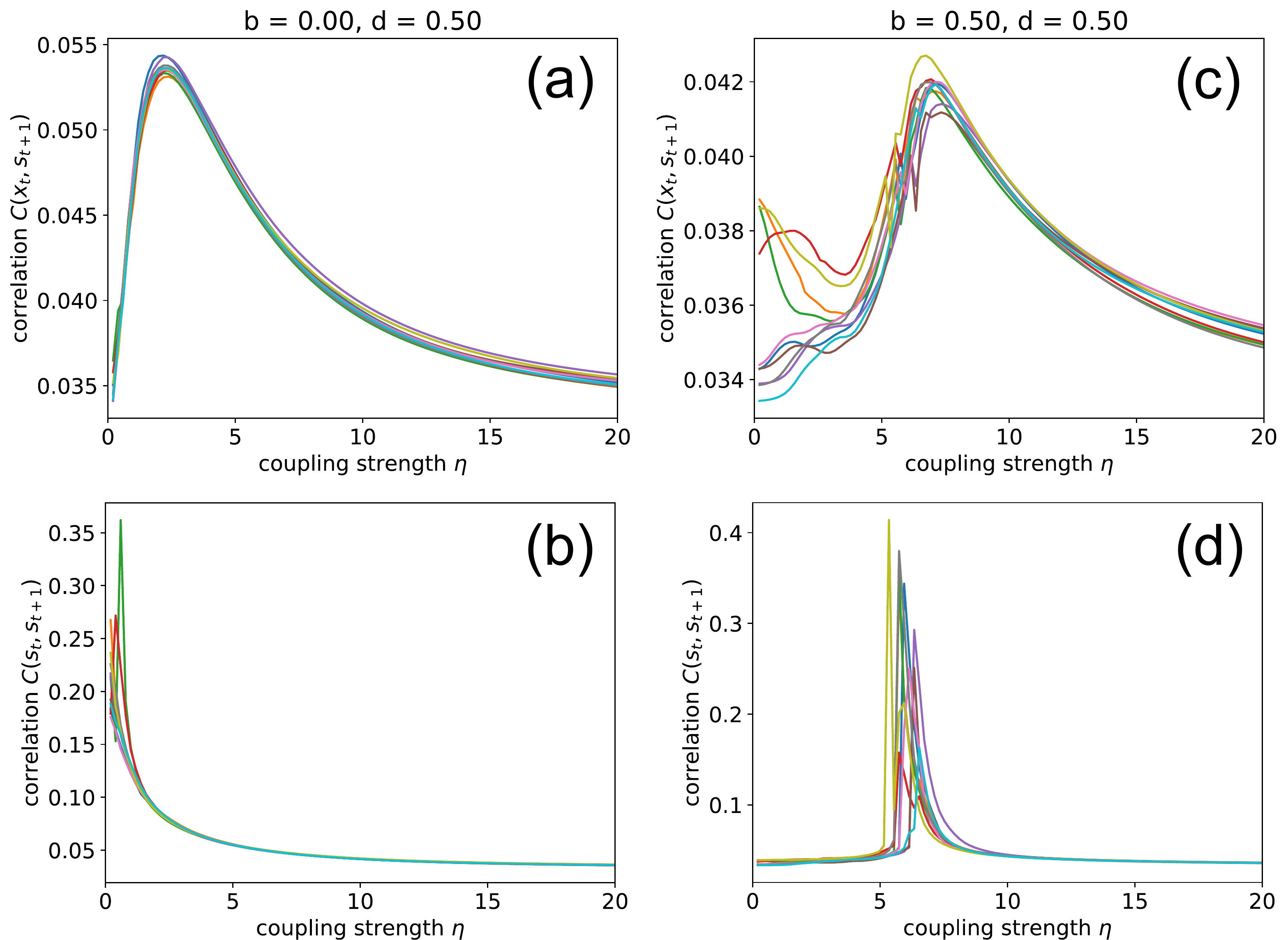}
\caption{Import resonance and recurrence resonance in RNNs. We compute the input-to-state  correlation $C(\mathbf{x}_t,\mathbf{s}_{t+1})$ (upper row) and the state-to-state correlation $C(\mathbf{s}_t,\mathbf{s}_{t+1})$ (lower row) for RNNs in the chaotic regime (left column) and in the fixed point regime (right column), as the coupling strength to the random (noise) input $\mathbf{x}_t$ is gradually increased from zero to 20. The computation has been repeated for 10 different realizations (colors) of RNNs with the given control parameters $b$ (balance) and $d$ (density). We find the phenomenon of import resonance (a,c) in both dynamical regimes, and the phenomenon of recurrence resonance (d) in the fixed point regime.} 
\label{figure_5}
\end{figure}

When in the fixed point regime (Fig. \ref{figure_5} d), we find that the dependence of the state-to-state correlation $C(\mathbf{s}_t,\mathbf{s}_{t+1})$ on the coupling strength $\eta$ has the shape of a 'resonance peak'. Since $\eta$ effectively controls the amplitude of 'noise' (used by us as pseudo input) added to the system, this corresponds to the phenomenon of 'Recurrence Resonance' (RR), which we have previously found in three-neuron motifs \cite{krauss2019recurrence}: At small noise levels $\eta$, the system is stuck in the fixed point attractor, but adding an optimal amount of noise (so that $C(\mathbf{s}_t,\mathbf{s}_{t+1})$ becomes maximal) is freeing the system from this attractor and thus makes recurrent information 'flux' possible, even in the fixed point regime. Adding too much noise is however counter-productive and leads to a decrease of $C(\mathbf{s}_t,\mathbf{s}_{t+1})$, as the system dynamics then becomes dominated by noise.
We do not observe recurrence resonance in the chaotic regime (Fig. \ref{figure_5} b), because recurrent information flux is possible there from the beginning.

Interestingly, we find very pronounced resonance-like curves also in the dependence of the input-to-state correlation $C(\mathbf{x}_t,\mathbf{s}_{t+1})$ on the coupling strength $\eta$, both in the chaotic (Fig. \ref{figure_5} a) and in the fixed point regime (Fig. \ref{figure_5} c). Since $C(\mathbf{x}_t,\mathbf{s}_{t+1})$ is a measure of information import, we call this novel phenomenon 'Import Resonance' (IR).

\subsection*{Network driven by continuous sinusoidal input}

\begin{figure}[ht!]
\centering
\includegraphics[width=1.0\linewidth]{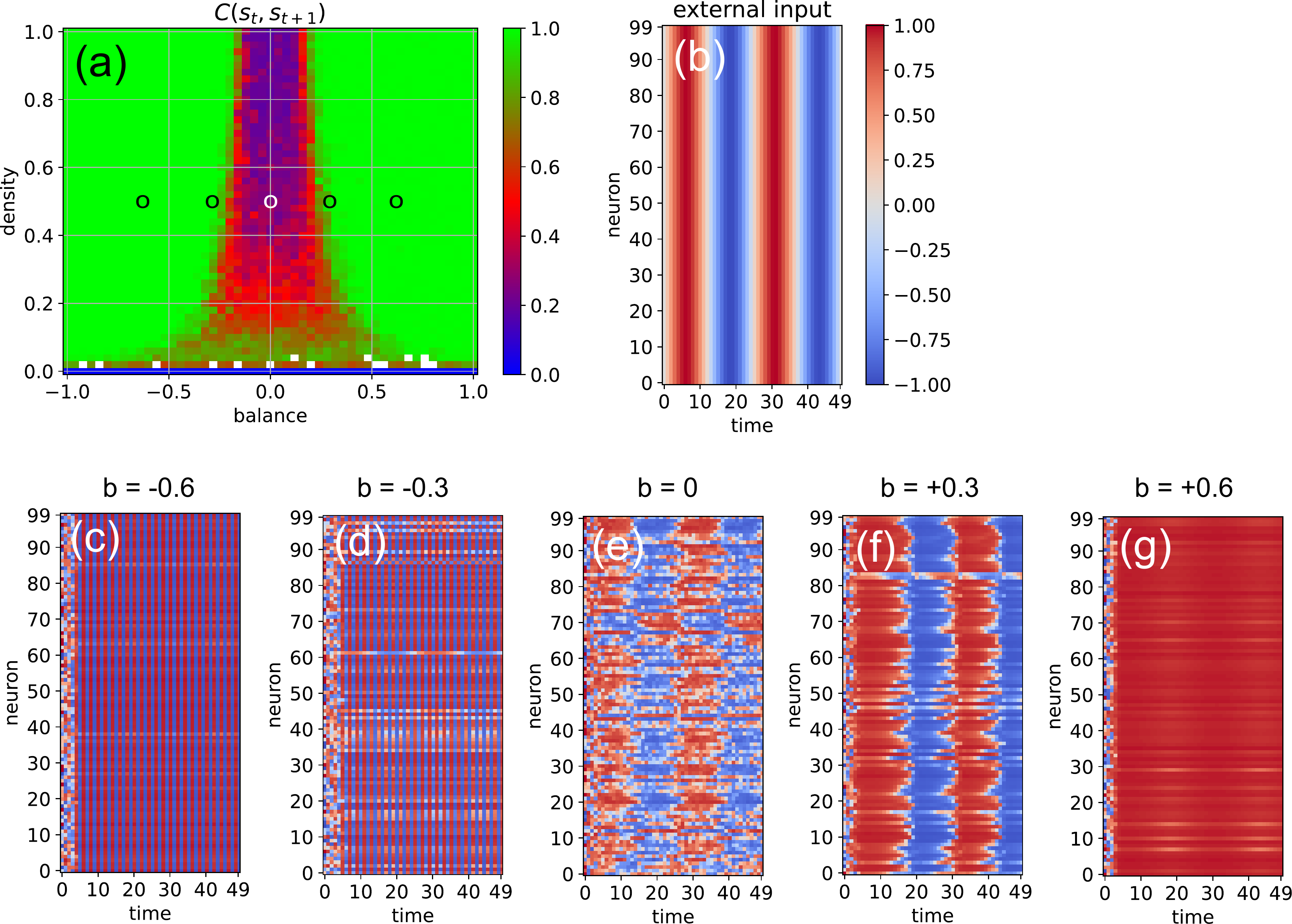}
\caption{Effect of a 'sinusoidal' input (b) on the activations of the RNN neurons (c-g) at five different points in the system's dynamic phase space (a). For all cases c-g, the density parameter is $d=0.5$, while the balance increases from -0.6 to +0.6. Only for balances sufficiently close to zero (d,e,f) the input is able to affect the system state.} 
\label{figure_6}
\end{figure}

Finally, we investigate the ability of the system to import more regular input signals with  built-in temporal correlations, as well as inputs that are identical for all neurons. For this purpose, we feed all neurons with the same sinusoidal input signal, using an amplitude of $a_{sin}=1$, an oscillation period of $T_{sin}=25$ time steps, and a coupling strength of $\eta=2$ (Fig. \ref{figure_6}).The density parameter is again fixed at $d=0.5$, while the balance increases from $b=-0.6$ to $b=+0.6$ in five steps. We find that the input signal does not affect the evolution of neural states when the system is too far in the oscillatory phase or too far in the fixed point phase (c,g). Only systems where excitatory and inhibitory connections are approximately balanced are capable of information import (d-f).
For $b=-0.3$ (d), most of the neurons are still part of the periodic attractor, but a small sub-population of neurons is taking up the external input signal (d). Interestingly, the system state is reflecting the periodic input signal even in the middle of the chaotic phase (e). 


\section*{Discussion}

In this study, we investigated RNNs with random weight matrices, their ability to import and process information, and how both abilities depend on the \emph{density of non-zero weights} and on the \emph{balance of excitatory and inhibitory connections}, as introduced in previous studies \cite{krauss2019weight, krauss2019analysis}.

It turned out that RNNs are simultaneously capable of both information import and information processing only in the low-density, i.e. sparse, part of the classical edge of chaos. Remarkably, this region of the phase space corresponds to the connectivity statistics known from the brain, in particular the cerebral cortex \cite{song2005highly, sporns2011non, miner2016plasticity}. In line with previous findings, i.e. that sparsity prevents RNNs from overfitting \cite{narang2017exploring, gerum2020sparsity} and is optimal for information storage \cite{brunel2016cortical}, we therefore hypothesize that cortical connectivity is optimized for both information import and processing. In addition, it seems plausible that there might be distinct networks in the brain that are either specialized to import and to represent information, or to process information and perform computations.

Furthermore, we found a completely new resonance phenomenon which we call \emph{import resonance}, showing that the correlation or mutual information between input and the subsequent network state depends on certain control parameters (such as coupling strength) in a peak-like way. Resonance phenomena are ubiquitous in biological \cite{mcdonnell2009stochastic} and artificial neural networks \cite{ikemoto2018noise, krauss2019recurrence, bonsel2021control} and have been shown to play a crucial role for neural information processing \cite{moss2004stochastic, krauss2018cross, schilling2020intrinsic}. In particular with respect to the auditory system, it has been argued that resonance phenomena like stochastic resonance are actively exploited by the brain to maintain optimal information processing \cite{krauss2016stochastic, krauss2017adaptive, krauss2018cross, schilling2021stochastic}. For instance, in a theoretical study it could be demonstrated that stochastic resonance improves speech recognition in an artificial neural network as a model of the auditory pathway \cite{schilling2020intrinsic}. Very recently, we were even able to show that stochastic resonance, induced by simulated transient hearing loss, improves auditory sensitivity beyond the absolute threshold of hearing \cite{krauss2021simulated}. The extraordinary importance of resonance phenomena for neural information processing indicates that the brain, or at least certain parts of the brain, do also actively exploit other kinds of resonance phenomena besides classical stochastic resonance. Whereas stochastic resonance is suited to enhance the detection of weak signals from the environment in sensory brain systems \cite{krauss2017adaptive}, we speculate that parts of the brain dealing with sensory integration and perception might exploit import resonance, while structures dedicated to transient information storage (short-term memory) \cite{ichikawa2020short} and processing might benefit from recurrence resonance \cite{krauss2019recurrence}. Similarly, the brain's action and motor control systems might also benefit from a hypothetical phenomenon of \emph{export resonance}, i.e. the maximization of correlation or mutual information between a given network state and a certain, subsequent readout layer.

Finally, our finding that both, correlation- and entropy-based measures of information import and processing yield almost identical phase diagrams (Fig. \ref{figure_3}, compare a, b with c, d), is in line with previously published results, i.e. that mutual information between sensor input and output can be replaced by the auto-correlation of the sensor output in the context of stochastic resonance (SR) \cite{krauss2017adaptive}. However, in this study we find that the equivalence of measures based on correlations and mutual information even extends to the phenomena of recurrence resonance (RR) \cite{krauss2019recurrence} and import resonance (IR), thereby bridging the conceptual gap (as described in \cite{mediano2021integrated}) between the information-processing perspective and the dynamical systems perspective on complex systems.


\section*{Methods}

\subsection*{Weight matrices with pre-defined statistics} 

\noindent We consider a system of $N_{neu}$ neurons without biases, which are mutually connected according to a weight matrix $\left\{w_{mk}\right\}$, where $w_{mk}$ denotes the connection strength from neuron $k$ to neuron $m$. The weight matrix is random, but controlled by three statistical parameters, namely the {\em 'density'} $d$ of non-zero connections, the excitatory/inhibitory {\em 'balance'} $b$, and the {\em 'width'} $w$ of the Gaussian distribution of weight magnitudes. The density ranges from $d=0$ (isolated neurons) to $d=1$ (fully connected network), and the balance from $b=-1$ (purely inhibitory connections) to $b=+1$ (purely excitatory connections). The value of $b=0$ corresponds to a perfectly balanced system.

\vspace{0.3cm}
\noindent In order to construct a weight matrix with given parameters $(b,d,w)$, we first generate a matrix $M^{(magn)}_{mn}$ of weight magnitudes, by drawing the $N_{neu}^2$ matrix elements independently from a zero-mean normal distribution with standard deviation $w$ and then taking the absolute value. Next, we generate a random binary matrix $B^{(nonz)}_{mn} \in \{0,1\}$, where the probability of a matrix element being $1$ is given by the density $d$, i.e. $p_1 = d$. Next, we generate another random binary matrix $B^{(sign)}_{mn} \in \{-1,+1\}$, where the probability of a matrix element being $+1$ is given by $p_{+1} = (1+b)/2$ where $b$ is the balance. Finally, the weight matrix is constructed by elementwise multiplication $w_{mn}=M^{(magn)}_{mn}\cdot B^{(nonz)}_{mn}\cdot B^{(sign)}_{mn}$. Note that throughout this paper, the width parameter is set to $w=0.5$.

\subsection*{Time evolution of system state} 

\vspace{0.3cm}
\noindent The momentary state of the RNN is given by the vector $\mathbf{s}(t)=\left\{s_m(t)\right\}$, where the component $s_m(t) \in \left[-1,+1\right]$ is the activation of neuron $m$ at time $t$. The initial state $\mathbf{s}(t\!=\!0)$ is set by assigning to the neurons statistically independent, normally distributed random numbers with zero mean and a standard deviation of $\sigma_{ini}=1$.

\vspace{0.3cm}
\noindent We then compute the next state vector by simultaneously updating each neuron $m$ according to
\begin{equation}
s_m(t+1) = \frac{2}{\pi} \; \arctan \left( 
\eta\;x_m(t) + \sum_{k=1}^N\; w_{mk}\; s_k(t) 
\right).
\end{equation}
Here, $x_m(t)$ are the external inputs of the RNN and $\eta$ is a global {\em 'coupling strength'}. Note that the input time series $x_m(t)$ can, but must not be different for each neuron. In one type of experiment, we set the $x_m(t)$ to independent, normally distributed random signals with zero mean and unit variance. In another experiment, we set all $x_m(t)$ to the same oscillatory signal $x(t) = a_{sin}\cdot\sin{\left(2\pi t/T_{sin}\right)}$.

\vspace{0.3cm}
\noindent After simulating the sequence of system states for $N_{stp}=1000$ time steps, we analyze the properties of the state sequence (see below). For this evaluation, we disregard the first $N_{tra}=100$ time steps, in which the system may still be in a transitory state that depends strongly on the initial condition. The simulations are repeated $N_{run}=10$ times for each set of control parameters ($b,d,\eta$).

\subsection*{Root-mean-squared pairwise correlation $C(\mathbf{u}_t,\mathbf{v}_{t+1}$)}

\vspace{0.3cm}
\noindent Consider a vector $\mathbf{u}(t)$ in $M$ dimensions and a vector $\mathbf{v}(t)$ in $N$ dimensions, both defined at discrete time steps $t$. The components of the vectors are denoted as $u_m(t)$ and $v_n(t)$. In order to characterize the correlations between the two time-dependent vectors by a single scalar quantity $C(\mathbf{u}_t,\mathbf{v}_{t+1})$, we proceed as follows: 

\vspace{0.3cm}
\noindent First, we compute for each vector component $m$ the temporal mean,
\begin{equation}
\mu_{um} = \left\langle u_m(t) \right\rangle_t
\end{equation}
and the corresponding standard deviation
\begin{equation}
\sigma_{um} = \sqrt{ \left\langle \left( u_m(t)-\mu_{um} \right)^2 \right\rangle_t }.
\end{equation}

\vspace{0.3cm}
\noindent Based on this, we compute the $M\!\times\!N$ pairwise (Pearson) correlation matrix,
\begin{equation}
C^{(uv)}_{mn} =
\frac{\left\langle\;
\left( u_m(t)-\mu_{um}  \right)\cdot \left( v_n(t\!+\!1)-\mu_{vn}  \right)
\;\right\rangle_t}{\sigma_{um} \; \sigma_{vn}},
\end{equation}
defining $C^{(uv)}_{mn}=0$ whenever $\sigma_{um}=0$ or $\sigma_{vn}=0$.

\vspace{0.3cm}
\noindent Finally we compute the root-mean-squared average of this matrix,
\begin{equation}
C(\mathbf{u}_t,\mathbf{v}_{t+1}) = \mbox{RMS}\left\{ C^{(uv)}_{mn} \right\}_{mn} =
\sqrt{\frac{1}{MN}\sum_{m=1}^M \sum_{n=1}^N \;\left| C^{(uv)}_{mn}\right|^2}
\end{equation}

\vspace{0.3cm}
\noindent This measure is applied in the present paper to quantify the correlations $C(\mathbf{s}_t,\mathbf{s}_{t+1})$ between subsequent RNN states, as well as the correlations $C(\mathbf{x}_t,\mathbf{s}_{t+1})$ between the momentary input and the subsequent RNN state.

\subsection*{Mean pairwise mutual information $I(\mathbf{u}_t,\mathbf{v}_{t+1}$)} 

\vspace{0.3cm}
\noindent In addition to the linear correlations, we consider the mutual information between the two vectors $\mathbf{u}(t)$ and $\mathbf{v}(t)$, in order to capture also possible non-linear dependencies. However, since the full computation of this quantity is computationally extremely demanding, we binarize the continuous vector components and then consider only the pairwise mutual information between these binarized components.  

\vspace{0.3cm}
\noindent For the binarization, we first subtract the mean values from each of the components,
\begin{equation}
u_m(t) \;\longrightarrow\; \Delta u_m(t) = u_m(t) - \mu_{um}.
\end{equation}

\vspace{0.3cm}
\noindent We then map the continuous signals $\Delta u_m(t) \in \left[-\infty,+\infty\right]$ onto two-valued bits $\hat{u}_m(t) \in \left\{ 0,1 \right\}$ by defining
$\hat{u}_m(t)\!=\!0$ if $\Delta u_m(t)<0$ and $\hat{u}_m(t)\!=\!1$ if $\Delta u_m(t)>0$. In the case of a tie, $\Delta u_m(t)=0$, we set $\hat{u}_m(t)\!=\!0$ with a probability of $1/2$. 

\vspace{0.3cm}
\noindent We next compute the pairwise joint probabilities $P(\hat{u}_m,\hat{v}_n)$ by counting how often each of the four possible bit combinations occurs during all available time steps. From that we also obtain the marginal probabilities $P(\hat{u}_m)$ and $P(\hat{v}_n)$.

\vspace{0.3cm}
\noindent The matrix of pairwise mutual information is then defined as
\begin{equation}
I^{(uv)}_{mn} = \sum_{\hat{u}_m=0,1} \;\sum_{\hat{v}_n=0,1}\;
P(\hat{u}_m,\hat{v}_n)\;
\log{
\left( 
\frac{P(\hat{u}_m,\hat{v}_n)}
{P(\hat{u}_m)\cdot P(\hat{v}_n)}
\right)},
\end{equation}
defining all terms as zero where $P(\hat{u}_m)\!=\!0$ or $P(\hat{v}_n)\!=\!0$.

\vspace{0.3cm}
\noindent Finally we compute the mean over all matrix elements (each ranging between 0 and 1 bit),
\begin{equation}
I(\mathbf{u}_t,\mathbf{v}_{t+1}) = \mbox{MEAN}\left\{ I^{(uv)}_{mn} \right\}_{mn} =
\frac{1}{MN}\sum_{m=1}^M \sum_{n=1}^N \; I^{(uv)}_{mn}
\end{equation}

\vspace{0.3cm}
\noindent This measure is applied in the present paper to quantify the mutual information $I(\mathbf{s}_t,\mathbf{s}_{t+1})$ between subsequent RNN states, as well as the mutual information $I(\mathbf{x}_t,\mathbf{s}_{t+1})$ between the momentary input and the subsequent RNN state.


\bibliographystyle{unsrt}
\bibliography{references}


\section*{Acknowledgements}
This work was funded by the Deutsche Forschungsgemeinschaft (DFG, German Research Foundation): grant KR\,5148/2-1 (project number 436456810) to PK.

\section*{Author contributions}
CM and PK have contributed equally to this paper.

\section*{Additional information}

\noindent{\bf Competing financial interests}
The authors declare no competing interests.  \vspace{0.5cm}

\noindent{\bf Data availability statement}
Data and analysis programs will be made available upon reasonable request.
\vspace{0.5cm}

\clearpage

\clearpage

\clearpage

\clearpage

\clearpage

\clearpage

\end{document}